\documentclass[11pt]{article}

\usepackage{fullpage}
\usepackage{amssymb}
\newcommand{\qed}{\hfill{$\rule{6pt}{6pt}$}} 
\newenvironment{proof}{\noindent{\bf Proof}:}{\qed} 
\newtheorem{theorem}{Theorem} 
\newtheorem{definition}{Definition} 

\newcommand{\POVM}{\mathsf{POVM}}
\newcommand{\defeq}{\stackrel{\Delta}{=}}

\newcommand{\Tr}{\mbox{{\rm Tr} }}

\newcommand{\lone}[1]{\| #1 \|_1}
\newcommand{\trace}[1]{\| #1 \|_t}

\title{\bf On divergence, relative entropy and the substate property}

\author{
Rahul Jain \\
U.C. Berkeley \\
{\sf rahulj@cs.berkeley.edu}
\thanks{
This work was supported by an Army Research Office (ARO), North
California,  grant number DAAD 19-03-1-00082.
}
\and 
Jaikumar Radhakrishnan \\
Tata Institute of Fundamental \\
Research, Mumbai, India \\
and \\
Toyota Technological Institute \\
Chicago, USA. \\
{jaikumar@tti-c.org}
\and
Pranab Sen \\
NEC Labs, Princeton \\
{pranab@nec-labs.com}
}

\begin{document}
\maketitle

\begin{abstract}
In this article we study relationship between three measures of
distinguishability of quantum states called as {\em
divergence}~\cite{jain:substate}, {\em
relative entropy} and the {\em substate property}~\cite{jain:substate}.  
\end{abstract}

\section{Introduction}
We consider three measures of distinguishability between quantum
states and show various relationships between them. The first measure
that we consider is called {\em divergence}. It was first considered
in~\cite{jain:substate} and is defined as follows:
\begin{definition}[Divergence]
Let $\rho, \sigma$ be two quantum states. Let $M$ be a {\em Positive
operator-valued measurement}($\POVM$). Then divergence between them
denoted $D(\rho | \sigma)$, is defined as, 
$$D(\rho | \sigma) \triangleq \max_{\mathsf{M:POVM}} \Tr M \rho \log \frac{\Tr M \rho} {\Tr M \sigma}$$
\end{definition}

The second is the well known measure called the {\em relative
entropy} also known as {\em Kullberg-Liebeck divergence}
(\cite{nielsen:quant}). It is defined as follows:
\begin{definition}[Relative entropy]
Let  $\rho, \sigma$ be quantum states. Then relative entropy
between them, denoted $S(\rho|\sigma)$ is defined as,
$$S(\rho|\sigma) \triangleq \Tr (\rho \log \rho - \rho \log \sigma)$$
\end{definition}

The third measure that we consider we call the {\em substate property}. It was
also first considered in~\cite{jain:substate}. It is defined as follows:
\begin{definition}[Substate property]
Two states $\rho$ and $\sigma$ are said to have the $k$-substate
property if,
$ \forall r >1, \exists \rho_r$ such that 

$$ \trace{\rho-\rho_r} \leq 2/\sqrt{r}  \mbox{   and   }  
\sigma - (1-\frac{1}{r}) \frac{\rho_r}{2^{rk}} \geq 0 $$
\end{definition}

\section{Results in this article}
The following theorem is a compilation of all the results in this article.
\begin{theorem}
\label{thm:divrelsub}
Let $\rho$ and $\sigma$ be two quantum states in $\mathbb{C}^n$.  Then,
\begin{enumerate}

\item
$D(\rho | \sigma) \leq  S(\rho|\sigma) + 1 $. This is not new and was shown
in~\cite{jain:substate}. We present a proof for completeness.

\item 
Given classical distributions $P$ and $Q$ on $[n]$, $ S(P|Q) \leq D(P | Q)
(n - 1). $

\item 
$  S(\rho|\sigma) \leq   D(\rho | \sigma) (n - 1) + \log n $.

\item
There exists classical distributions $P$ and $Q$ on $[n]$ such that,
$$  S(P|Q) >   (D(P | Q)/2 - 1) (n -2) - 1.  $$

\item 
{\bf (Substate theorem)} $\rho$ and $\sigma$ have the
$(8D(\rho|\sigma) + 14)$-substate property. This is not new and was
shown in~\cite{jain:substate}. Please refer to~\cite{jain:substate}
for a proof.

\item 
{\bf (Converse of substate theorem for classical distributions)}
If distributions $P$ and $Q$ have the $k$-substate property then
$D(P|Q) \leq 2k +2$.

\item If the following {\em strong $k$-substate property} holds, i.e.
$ \sigma - \frac{\rho}{2^{k}} \geq 0$, then $S(\rho|\sigma) \leq k$.

\item There exists
a POVM such that, if $P$ and $Q$ are resulting classical distributions
then, $$ S(P|Q) \geq \frac{S(\rho|\sigma) - \log n}{n-1} - 1.$$
\end{enumerate}
\end{theorem}
\begin{proof} 
\begin{enumerate}
\item 
Let $M$ be the POVM that achieves $D(\rho|\sigma)$. Let $\Tr M \rho
\defeq p$ and $\Tr M \sigma \defeq q$. 
\begin{eqnarray*}
S(\rho | \sigma )
  & \geq & p \log \frac{p}{q} +
           (1 - p) \log \frac{(1 - p)}{(1 - q)} \\
  &   >  & p \log \frac{p}{q} +
           (1 - p) \log \frac{1}{(1 - q)} - 1 \\
  & \geq & p \log \frac{p}{q} - 1 \\
  &   =  & D(\rho | \sigma) - 1.
\end{eqnarray*}
The first inequality follows from the {\em Lindblad-Uhlmann monotonicity}
of relative entropy~\cite{nielsen:quant}
and the second
inequality follows because
$(1 - p) \log (1 - p) \geq (- \log e)/e > -1$,
for $0 \leq p \leq 1$. 
\qed\\
\item
Define $x_i = \log (p_i/q_i)$. 
We can assume without loss of generality, by perturbing $Q$ slightly,
that the values $x_i$ are distinct for distinct $i$.
Let $S' = \{i : x_i > 0\}$. 
Let $D(P|Q)  = k$. Let 
$$ \forall \mbox{positive real } l, 
S_l = \{ i \in [n] : x_i \geq l \}.$$ 
Therefore,
\begin{eqnarray*}
k &\geq& \Pr_P[S_l] \log \frac{\Pr_P[S_l]}{\Pr_Q[S_l]} 
\geq \Pr_P[S_l] l
\\
\Rightarrow \Pr_P[S_l] &\leq&  k/l
\end{eqnarray*}
Assume without loss of generality that $x_1 < x_2 < \cdots < x_n$.
Then if $x_i > 0$, $\Pr_P[S_{x_i}] \leq k/x_i$.
Since $S(P|Q) \leq \sum_{i \in S'} p_i x_i$, the upper bound on 
$S(P|Q)$
is maximized when $S' = \{2, \ldots, n\}$, $p_n = k/x_n$, 
$p_i = k (1/x_i - 1/x_{i+1})$ for all $i \in \{2, \ldots, n - 1\}$,
$p_n = k/x_n$,
and $p_1 = 1 - \sum_{i=2}^n p_i$. Then,
\begin{eqnarray*}
S(P|Q) &\leq& \sum_{i=2}^{n} p_i x_i 
\\
&=& k \sum_{i=2}^{n-1} x_i (1/x_i - 1/x_{i+1}) + k
\\
&=& k \sum_{i=2}^{n-1} \frac{x_{i+1} - x_i}{x_{i+1}} + k
\\
&\leq& k \sum_{i=2}^{n-1} 1 + k
\\
&=& k (n - 1).
\end{eqnarray*}
\qed
\item Let us measure $\rho$ and $\sigma$ in the eigenbasis of
$\sigma$. We get two distributions: $P_{\rho}$ and $P_{\sigma}$.  Let
$D(P_{\rho} | P_{\sigma}) = k$.  From Part 2, it follows,
\begin{eqnarray*}
 k (n - 1) =  D(P_{\rho} | P_{\sigma}) (n - 1) & \geq & S(P_{\rho} | P_{\sigma})  
\\
	& = & \Tr P_{\rho} \log P_{\rho} - \Tr P_{\rho} \log P_{\sigma} 
\\
                             & \geq & - \log n - \Tr P_{\rho} \log P_{\sigma}
\\
                              & = &  - \log n - \Tr \rho \log \sigma
\\
                            & = &  - \log n + S(\rho | \sigma) - \Tr \rho \log \rho
\\
                          & \geq &  - \log n + S(\rho | \sigma)
\end{eqnarray*}
The second equality above holds since the measurement was in the
eigenbasis of $\sigma$.

Thus $$  S(\rho | \sigma) \leq D(P | Q) (n - 1) + \log n$$ \qed  
\item
Fix $a > 1$, $k > 0$.
Let $p_1 = (a - 1)/a$,
$$\forall i \in \{2, \ldots, n-1\} ~~~~~ p_i = (a - 1)/a^i,$$
and $p_n = 1/a^{n - 1}$.
Let 
$$\forall i \in \{2, \ldots, n\} ~~~~~ q_i = p_i/2^{k a^{i - 1}},$$
and $q_1 = 1 - \sum_{i=2}^n q_i$.
For any $r > 1$, consider 
$\tilde{P} = (p_1, \ldots, p_{\log_a r + 1}, 0, \ldots, 0)$ normalized
to make it a probability vector. It is easy to see that 
$\lone{P - \tilde{P}} \leq 2/r$ and 
$\frac{(r-1) \tilde{P}}{r 2^{rk}} \leq Q$. This shows that $P, Q$ 
satisfy the classical $k$-substate property, hence $D(P|Q) \leq 2(k+1)$.
 
Now,
\begin{eqnarray*}
S( P | Q) 
&  = & \sum_{i=1}^n p_i \log \frac{p_i}{q_i} \\
&\geq& p_1 \log p_1 + \sum_{i=2}^n p_i \log \frac{p_i}{q_i} \\
&  > & -1 + (n-2) \frac{k(a-1)}{a} + k \\
&  = & k (n - 1) - \frac{k(n-2)}{a} - 1.
\end{eqnarray*}
By choosing $a$ large enough, we can achieve 
$S(P|Q) > k (n - 2) - 1$. This shows that 
$S(P|Q) > (D(P|Q)/2 - 1)(n - 2) - 1$.
\item Proof skipped. Please see~\cite{jain:substate} for a
detailed proof.\qed 
\item Let $M$ be a POVM such that 
$$k_1 \triangleq D(P | Q) = \Tr M P \log \frac{\Tr M P}{\Tr M Q} $$
Let  $p \triangleq \Tr M P $ and $ q \triangleq \Tr M Q $. Therefore,
$$ k_1 = p \log \frac{p}{q} \Rightarrow q = \frac{p}{2^{k_1/p}}$$
Let $r= 2/p$. Since $P$ and $Q$ have the $k$-substate
property, let $P_r$ be the distribution such that,
\begin{equation}
\label{eq:sub1}
\trace{P-P_r} \leq \frac{2}{r} = p \mbox{ (holds for classical
distributions~\cite{jain:substate}) } 
\end{equation}
and 
\begin{equation}
\label{eq:sub2}
Q - (1-\frac{1}{r}) \frac{P_r}{2^{rk}} \geq 0 
\end{equation}
Let $p_r \triangleq \Tr M P_r $. From (\ref{eq:sub1}) it follows,
\begin{equation}
\label{eq:prp}
p_r \geq \frac{p}{2}  
\end{equation}
Also 
\begin{eqnarray*}
 \frac{p}{2^{k_1/p}}= q  & = &  \Tr M Q 
\\
& \geq &  (1-\frac{1}{r}) \Tr \frac{M P_r}{2^{rk}} ~~~\mbox{(from (\ref{eq:sub2}))}
\\
& = & (1- \frac{p}{2})  \frac{p_r}{2^{rk}}  ~~~~\mbox{(from definition)}
\\
& \geq & (\frac{1}{2}) \frac{p_r}{2^{rk}} ~~~~\mbox{(since} ~~p \leq
1 \mbox{)}
\\
& \geq & (\frac{1}{2}) \frac{p}{2^{rk + 1}}  ~~~~\mbox{(from (\ref{eq:prp}))}
\\
\Rightarrow 2^{rk+2} & \geq  &  2^{k_1/p}
\\
\Rightarrow rk+2 & \geq & k_1/p
\\
\Rightarrow p(rk+2) & \geq & k_1
\\
\Rightarrow prk+ 2 & \geq & k_1 ~~~~\mbox{(since} ~~p \leq
1 \mbox{)}
\\
\Rightarrow 2k+ 2 & \geq &  k_1 
\end{eqnarray*}
\qed \\
\noindent
{\bf Remark :} This proof does not work for the quantum case
because of $\sqrt(r)$ in the substate property. 
\item
\begin{eqnarray*}
S(\rho) &=& \Tr \rho \log \rho - \Tr \rho \log \sigma 
\\
 &\leq& \Tr \rho \log \rho - \Tr \rho \log \frac{\rho}{2^{k}}
\\
&=& k \Tr \rho = k
\end{eqnarray*}
The first inequality above follows from monotonicity of the
operator log function.
\qed 
\item
We know that there exists a POVM element $M$ such that,
$$D(\rho|\sigma) = \Tr M \rho \log \frac{\Tr M \rho}{\Tr M \sigma}$$
Let $p = \Tr M \rho$ and $q = \Tr M \sigma$. Let $P = (p, 1-p)$ and $Q
= (q, 1-q)$. 
Note that 
\begin{eqnarray*}
 S(P|Q) &=& p \log p/q + (1-p) \log (1-p)/(1-q)
\\
&\geq&  p \log p/q  - 1
\\
&=&  D(\rho|\sigma)  - 1
\end{eqnarray*}
From Part 2 it follows that 
\begin{eqnarray*}
S(\rho|\sigma) &\leq& D(\rho|\sigma) (n - 1) + \log n
\\
&\leq& (S(P|Q) + 1) (n - 1) + \log n
\\
\Rightarrow  S(P|Q) &\geq& \frac{S(\rho|\sigma) - \log n}{n-1} - 1
\end{eqnarray*}
\end{enumerate}
\end{proof}

\end{document}